\newif\ifabstract
\abstracttrue
 \abstractfalse 
\newif\iffull
\ifabstract \fullfalse \else \fulltrue \fi

\documentclass[11pt]{article}
\usepackage{graphicx}
\usepackage{url}
\usepackage{authblk}

\newcommand{\sectionref}[1]{Section \ref{section:#1}}

\newcommand{\sectionlabel}[1]{\label{section:#1}}
\newcommand{\sectionlabeled}[2]{\section{#2}{\sectionlabel{#1}}}
\newcommand{\subsectionlabeled}[2]{\subsection{#2}{\sectionlabel{#1}}}

\makeatletter
\def\figref{\@ifnextchar[{\@figref}{\@figref[]}}
\def\@figref[#1]#2{\mbox{Figure {\ref{figure:#2}#1}}}
\def\tabref{\@ifnextchar[{\@tabref}{\@tabref[]}}
\def\@tabref[#1]#2{\mbox{Table {\ref{table:#2}#1}}}
\makeatother

\newcommand{\figlabel}[1]{\label{figure:#1}}
\newcommand{\tablabel}[1]{\label{table:#1}}

\newcommand{\ie}{i.e.~\/}
\newcommand{\eg}{e.g.~\/}

\advance \oddsidemargin by -0.8in
\newcommand{\myparskip}{3pt}
\parskip \myparskip

\begin{document}

\title{Automatic Recognition of Public Transport Trips from Mobile Device Sensor Data and Transport Infrastructure Information}
\author{Mikko Rinne}
\author{Mehrdad Bagheri}
\author{Tuukka Tolvanen}
\affil{Aalto University School of Science\\Department of Computer Science\\Email: {\tt firstname.lastname@aalto.fi}.}

\begin{titlepage}
\maketitle

\thispagestyle{empty}

\begin{abstract}
Automatic detection of public transport (PT) usage has important applications for intelligent transport systems. It is crucial for understanding the commuting habits of passengers at large and over longer periods of time. It also enables compilation of door-to-door trip chains, which in turn can assist public transport providers in improved optimisation of their transport networks. In addition, predictions of future trips based on past activities can be used to assist passengers with targeted information.
This article documents a dataset compiled from a day of active commuting by a small group of people using different means of PT in the Helsinki region. Mobility data was collected by two means: (a) manually written details of each PT trip during the day, and (b) measurements using sensors of travellers' mobile devices. The manual log is used to cross-check and verify the results derived from automatic measurements. The mobile client application used for our data collection provides a fully automated measurement service and implements a set of algorithms for decreasing battery consumption.
The live locations of some of the public transport vehicles in the region were made available by the local transport provider and sampled with a 30-second interval. The stopping times of local trains at stations during the day were retrieved from the railway operator. The static timetable information of all the PT vehicles operating in the area is made available by the transport provider, and linked to our dataset. The challenge is to correctly detect as many manually logged trips as possible by using the automatically collected data. This paper includes an analysis of challenges due to missing or partially sampled information in the data, and initial results from automatic recognition using a set of algorithms. Improvement of correct recognitions is left as an ongoing challenge.
\end{abstract}

\end{titlepage}

\sectionlabeled{introduction}{Introduction}
Automatic detection of the door-to-door trip chains of people has a multitude of applications. For infrastructure planners and public transport providers knowledge of the true origins, destinations and volumes of commuters gives a much better understanding of requirements for the road and transport networks than disconnected counts of cars or people at points of observation of the current network by using loops, cameras, ticket systems or manual passenger counting campaigns.

To assist users of public transport with relevant information about opportunities and problems it is vital to be able to generate a prediction of the next destination, time of travel and the means of public transport the person is going to use. For many people the same trips are repeated with regular cycles, \ie daily, on (certain) weekdays, weekly or monthly. Such travellers can be proactively given targeted information about disruptions in road traffic or public transport lines, which they frequently use, at their personal times of regular usage. Real-time recognition of the current means of public transport and a prediction of the likely destinations can also be used to assist connecting passengers.

Multiple smartphone-assisted ways for automatic detection of public transport usage can be envisaged:
\begin{itemize}
\item \emph{Ticketing system:} If coupled with the payment of the trip in systems where both vehicle entries and exits are registered, precise and correct information about the public transport part of trips can be obtained. Requires integration with the fare payment system.
\item \emph{Radio beacon:} Some transport providers may provide vehicle-installed radio transmitters, \eg \emph{WiFi\footnote{\url{http://www.wi-fi.org/}}} or \emph{Bluetooth\footnote{\url{https://www.bluetooth.com/}}}, which can be detected and compared with a list of beacon identifiers to detect the proximity of a public transport vehicle.
\item \emph{Live positioning:} The live positions of public transport vehicles can be matched with the measured positions of the passenger, searching for a sequence of continuous matches with suitable accuracy.
\item \emph{Static timetable:} Processed information about a trip carried out by a person (start and end locations and times, route geometry) using a vehicle can be compared with the information in a static timetable.
\end{itemize}
In our test arrangement we did not have access to the ticketing system of the public transport provider. Additionally, the current policy in the Helsinki region does not require registration of vehicle exits, and one validation of a regional ticket allows up to 80 minutes of transfer without a new validation. Therefore ticket-based information would not have been accurate even if it were available. At the time of testing listed Wi-Fi beacons were only available on trams, which does not give an adequate coverage of the transport network. Live positioning of public transport vehicles was made available by the public transport provider for a part of their fleet. Static timetable information was available for all lines and departures.

The target was to create a dataset for testing and benchmarking algorithms for automatic recognition of public transportation trips. The dataset is composed of position and activity recognition samples of 8 researchers between 9 am and 4 pm EET+DST on August 26th 2016, manual bookkeeping of their trips and the related transport infrastructure data. Data collection for the limited period was pre-agreed with every campaign participant to enable publication of the dataset without privacy concerns. 

Seven participants executed as many public transportation trips as possible during the designated time, especially emphasising travel by subway, as it has been the most challenging transportation mode for automatic recognition. The eighth participant logged some private car trips to provide comparison data, which should not match with any public transportation.

Due to the exceptional amount of travel per person during one day this dataset cannot be used as a source for studying the regular travel habits of public transportation users. It also doesn't contain repeatedly visited locations such as homes or offices. The challenge is to correctly recognise as many trips listed in the manual log as possible by using the other forms of data available. The dataset consists of the following tables:
\begin{itemize}
\item \emph{Device data:} samples from mobile device sensors.
\item \emph{Filtered device data:} selection of the perceived activity and exclusion of data points at times of staying still using our algorithms.
\item \emph{Device models:} phone models used by the participants.
\item \emph{Manual log:} manual trip bookkeeping entries of participants.
\item \emph{Live position samples:} public transport fleet positions.
\item \emph{Static timetables:} public transport timetables valid on the date of the experiment.
\item \emph{Train stop times:} measured train stop time information for the date of the experiment.
\end{itemize}
The complete dataset is available\footnote{Static timetable data is referenced from the resources of the transport provider.} in \emph{github\footnote{\url{https://github.com/aalto-trafficsense/public-transport-dataset}}}. 
\sectionlabeled{background}{Background}
Transportation mode estimation and classification using mobile phone sensors has been discussed \eg in \cite{Feng2013,Hemminki2013,Shin2015}. Automatic recognition of bus trips using mobile phone sensors has earlier been addressed by the \emph{Live+Gov\footnote{\url{http://liveandgov.eu/}}} project. Their mobile client collected both hardware sensor (accelerometer, rotation vector, gyroscope, magnetic field) and processed activity detection data \cite{Hertmann2014}. A project proprietary human activity recognition classifier was built and trained. Public transport recognition was attempted using public transport infrastructure data similar to our study (static timetables and live locations of public transport vehicles). The client software also supported user marking of activity, but users were shown the currently detected activity, and consequently they have generally registered their activities only in cases, where the detected activity was incorrect. Service line detection probability of 37\% is reported \cite{Minnigh2015}. 

In one study bus passengers with mobile phones have been used as voluntary crowdsourced sensors for reporting the live locations of buses \cite{Corsar2013}. In the Live+Gov project traffic jams were detected from irregularities in public transport fleet location data \cite{Hertmann2014}. The specific context of parking has also been considered, estimating from mobile phone sensor data, whether a parking area has space available \cite{Nandugudi2014,Rinne2014b}.

The present effort belongs to the \emph{TrafficSense\footnote{\url{http://trafficsense.aalto.fi}}} project, a part of the \emph{Aalto Energy Efficiency Research Programme\footnote{\url{http://aef.aalto.fi/en/}}}. The project aims to save time and energy in traffic by understanding regular travel habits of individuals and proactively assisting them with information available in the internet. Sustainability and business model of the approach have been considered in \cite{Heiskala2016}. The current prototype service uses the methods outlined in this paper to discover the public transport usage of travellers and automatically and proactively relay targeted information about disruptions in the greater Helsinki area based on the individually detected lines and times of usage.

The current TrafficSense mobile client is coded in native Android\footnote{Android\textsuperscript{TM} is a trademark of Google Inc.}. It is available on \emph{Google Play\footnote{\url{https://play.google.com/store/apps/details?id=fi.aalto.trafficsense.trafficsense}}}, but currently restricted to Finland, because the public transport recognition and disruption information functions are only available locally. Example screenshots are shown in \figref{trafficsense-screenshots}. The main map view a) shows ranked regular destinations (purple 4 and 5), car, bus and train trips as well as the current location of the user. A train trip is selected with departure and arrival times indicated. The ``Edit/Confirm'' button can be used by the user to confirm or edit the activity and the line name (currently ``U''). The screenshot of the notification tray b) shows a public transport disruption bulletin.
\begin{figure}
\begin{center}
\includegraphics[width=0.8\columnwidth]{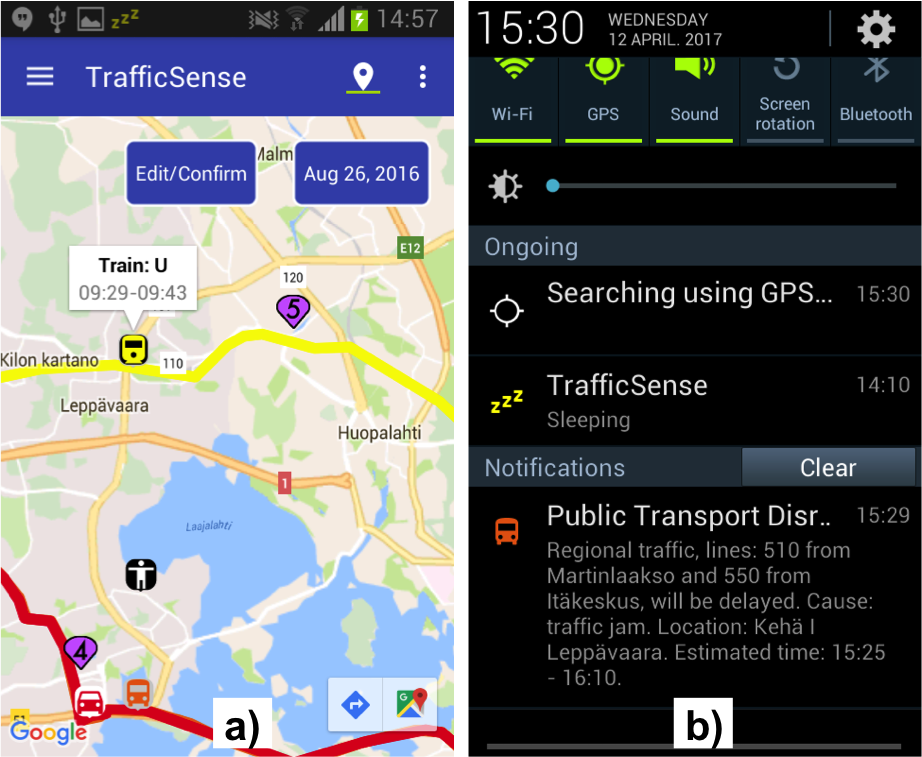}
\caption{Example screenshots from the TrafficSense client. Map data \copyright 2016 Google.}
\figlabel{trafficsense-screenshots}
\end{center}
\end{figure}

The types, formats and sources of open traffic information available were reviewed in \cite{Rinne2014a}. For information on public transport the resources from \emph{Helsinki Regional Transport\footnote{\url{https://www.hsl.fi/en}}} (HRT) were used. Their developer resources are currently provided and documented under the \emph{digitransit\footnote{\url{https://www.digitransit.fi/en/developers/}}} umbrella. The resources follow the \emph{SIRI} (Service Interface for Real-time Information \cite{Rinne2014a}) specifications\footnote{\url{http://user47094.vs.easily.co.uk/siri/}}. Live locations of the fleet were sampled from their live data API\footnote{At the time of the study, the data was available at \url{http://dev.hsl.fi/siriaccess/vm/json}. The address has been later replaced by \url{http://api.digitransit.fi/realtime/vehicle-positions/v1/siriaccess/vm/json}}. The static timetable data is in \emph{GTFS} (General Transit Feed Specification\footnote{\url{https://developers.google.com/transit/gtfs/}}) format as specified by Google Inc. 

Train stop time information is encoded as \emph{JSON} (Javascript Object Notation\footnote{\url{http://www.json.org/}} \cite{ECMA-4042013}). It has been fetched from the \emph{digitraffic\footnote{\url{http://www.liikennevirasto.fi/web/en/open-data/digitraffic\#.WOt_a1OGOHo}}} service\footnote{\url{http://rata.digitraffic.fi/api/v1/history?departure_date=2016-08-26}} operated by the \emph{Finnish Transport Agency\footnote{\url{http://www.liikennevirasto.fi/web/en}}} (FTA). All other sampled data in the repository is made available in \emph{CSV} (Comma-Separated Values \cite{Shafranovich2005}) format. The repository includes scripts for importing the CSV tables into a \emph{PostgreSQL\footnote{\url{https://www.postgresql.org/}}} database, which is used internally by the TrafficSense project.
\sectionlabeled{manual-log}{Manual bookkeeping by test participants}
All test participants manually documented the details of their trips during the day. The information provided for each trip leg is shown in \tabref{manual-log}. Because the timestamps were manually recorded, their values are approximate. In total 103 trips were recorded. 
\begin{table}[htp]
\footnotesize
\caption{Manually logged trip information (tz = timezone).}
\begin{center}
\begin{tabular}{|l|l|l|}
\hline
\textbf{Label} & \textbf{Type} & \textbf{Description}\\
\hline
\texttt{device\_id} & integer & Device identifier, aligned with \texttt{device\_id} in \sectionref{device-data}.\\
\hline
\texttt{st\_entrance} & string & Description of entrance to station building, if applicable.\\
& &  A map of the letters to mark entrances at each subway station\\
& & is provided in the repository.\\
\hline
\texttt{st\_entry\_time} & timestamp (no tz) & Station entry time, if applicable.\\
\hline
\texttt{line\_type} & string & \texttt{SUBWAY} / \texttt{BUS} / \texttt{TRAM} / \texttt{TRAIN} / \texttt{CAR}\\
\hline
\texttt{line\_name} & string & Identifier, \eg 7A, 102T, U, V.\\
\hline
\texttt{vehicle\_dep\_time} & timestamp (no tz) & Vehicle departure time.\\
\hline
\texttt{vehicle\_dep\_stop} & string & Description of the platform or other station where vehicle was boarded.\\
\hline
\texttt{vehicle\_arr\_time} & timestamp (no tz) & Vehicle stop time at the end of the trip.\\
\hline
\texttt{vehicle\_arr\_stop}  & string & Description of the platform or other station where the vehicle was exited.\\
\hline
\texttt{st\_exit\_location} & string & Description of exit to station building, if applicable. Subway exit letters\\
& & have been marked in the maps provided with the above footnote.\\
\hline
\texttt{st\_exit\_time} & timestamp (no tz) & Time of exiting station, if applicable.\\
\hline
\texttt{comments} & string & Freeform comments about the trip leg.\\
\hline
\end{tabular}
\end{center}
\tablabel{manual-log}
\end{table}
\sectionlabeled{device-data}{Mobile device measurements}
Mobile device samples were collected using the TrafficSense mobile client. The client uses the fused location provider and activity recognition of \emph{Google Play Services\footnote{\url{https://developers.google.com/android/guides/overview}}}. Both of them are \emph{virtual sensors} \cite{Kabadayi2006}, abstracting information from available hardware sensors\footnote{the availability of a particular hardware sensor may vary between different device models} into current best estimates of the location of the device and the activity the person carrying the device is currently engaged in. The sampled parameters contained in the dataset are listed in \tabref{device-data}. The coordinates are in \emph{WGS84\footnote{\url{http://gisgeography.com/wgs84-world-geodetic-system/}}} format. The table in the dataset contains 6,030 entries. It is formatted as CSV and sorted by \texttt{time} and \texttt{device\_id}.
\begin{table}[htp]
\footnotesize
\caption{Parameters sampled from mobile devices (tz = timezone).}
\begin{center}
\begin{tabular}{|l|l|l|}
\hline
\textbf{Label} & \textbf{Type} & \textbf{Description}\\
\hline
\texttt{time} & timestamp (no tz) & From the clock of the mobile device.\\
\hline
\texttt{device\_id} & integer & Stable identifier for the device.\\
\hline
\texttt{lat} & double & Latitude, in WGS84. \\
\hline
\texttt{lng} & double & Longitude, in WGS84.\\
\hline
\texttt{accuracy} & double & Radius, in meters, estimated by the \emph{fused location provider}\\
& & of the mobile device.\\
\hline
\texttt{activity\_1} & enum & Activity with highest confidence, provided by \emph{activity recognition} of\\
& & \emph{Google Play Services}. Values: \texttt{IN\_VEHICLE}, \texttt{ON\_BICYCLE}, \texttt{RUNNING}, \texttt{STILL},\\
& & \texttt{TILTING}, \texttt{UNKNOWN}, \texttt{WALKING}.\\
\hline
\texttt{activity\_1\_conf} & integer & Percentage of recognition certainty, 100  = best confidence.\\
\hline
\texttt{activity\_2} & enum & Value of second-highest confidence activity.\\
\hline
\texttt{activity\_2\_conf} & integer & Percentage of recognition certainty.\\
\hline
\texttt{activity\_3} & enum & Value of third-highest confidence activity.\\
\hline
\texttt{activity\_3\_conf} & integer & Percentage of recognition certainty.\\
\hline
\end{tabular}
\end{center}
\tablabel{device-data}
\end{table}
\subsectionlabeled{client-filtering}{Mobile client filtering algorithms}
The client alternates between \texttt{ACTIVE} and \texttt{SLEEP} states as shown in \figref{active-sleep-state-diagram}. The current default value for the sleep timer is 40 seconds. If the detected position changes by a distance longer than the accuracy of the position fix during a period of the perceived activity indicating \texttt{STILL}, the timer is restarted. This rule aims to prevent erroneous transitions to \texttt{SLEEP} state while the device is moving, but activity detection perceives it as being still. Such situations typically occur during smooth trips on rails, \ie trains, trams and subways.
\begin{figure}
\begin{center}
\includegraphics[width=0.7\columnwidth]{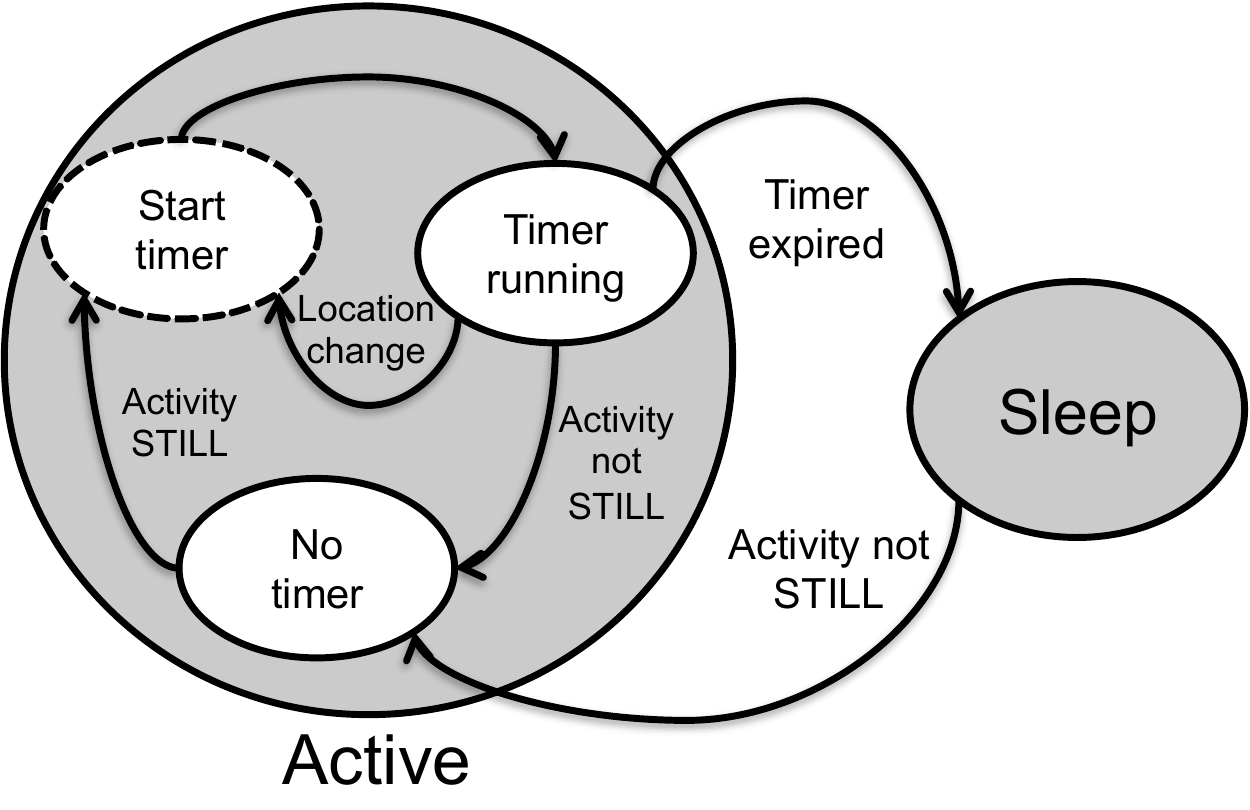}
\caption{State transitions between mobile client \texttt{ACTIVE} and \texttt{SLEEP} states.}
\figlabel{active-sleep-state-diagram}
\end{center}
\end{figure}

In \texttt{ACTIVE} state position is requested from the \emph{fused location provider\footnote{https://developers.google.com/android/reference/com/google/android/gms/location/FusedLocationProviderApi}} provided by Google Play Services with \emph{high accuracy} and a 10 second interval. In \texttt{SLEEP} state position requests are dropped to \emph{no power} priority, which means that position fixes are passed to our client only if requested by another application. Activity reports are always requested with a 10 second interval, but as a form of device power saving, during \texttt{STILL} detections activity recognition interval has been observed to increase up to 180 seconds. As a result, sometimes the client may need up to $\approx 200 \mathrm{s}$ of movement to wake up from \texttt{SLEEP} state.

In our data format each accepted position record is coupled with the latest activity information. The timestamp of the entry is the timestamp of the position, not the activity detection. Therefore the same detected activity may repeat over multiple points. The received position fixes (`points') are filtered as follows (\figref{sensor-data-filter-algorithm}):
\begin{itemize}
\item If 60 minutes have passed since the last accepted point, any point is accepted (to provide a ``ping'' effect and record that the client was running).
\item Accuracy\footnote{Estimated by the \emph{fused location provider}.} must be better than 1000m.
\item If \texttt{activity != last queued activity} and \texttt{activity} is `good'\footnote{not \texttt{UNKNOWN} or \texttt{TILTING}}, the point is accepted.
\item If \texttt{(activity == last queued activity) and (distance to last accepted point > accuracy)}, the point is accepted.
\end{itemize}
The dataset contains about 30\% of the theoretical maximum\footnote{one point every 10 seconds from every terminal} number of points.
\begin{figure}
\begin{center}
\includegraphics[width=1.0\columnwidth]{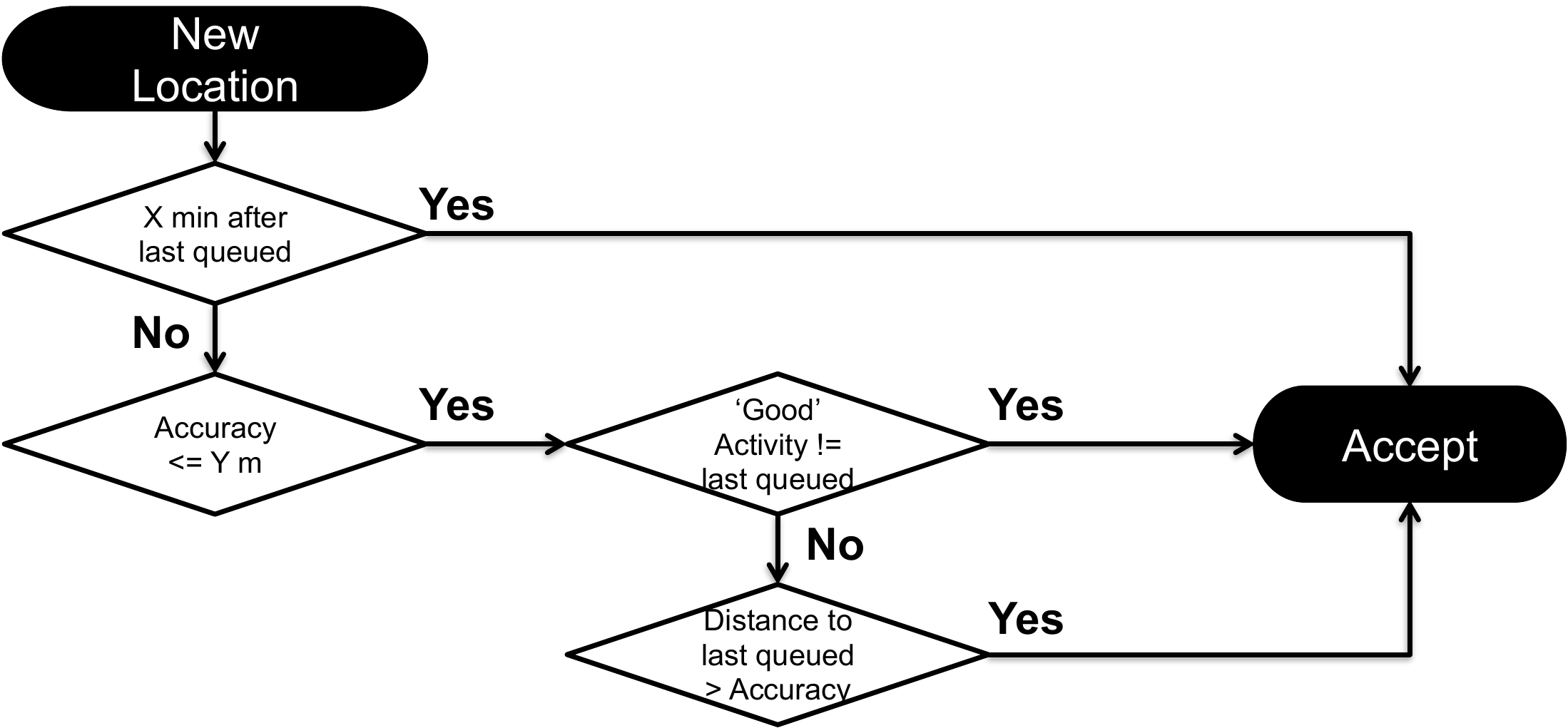}
\caption{Filtering algorithm for incoming position fixes.}
\figlabel{sensor-data-filter-algorithm}
\end{center}
\end{figure}

The fused location provider used by the mobile client combines data from satellite, Wi-Fi and cellular positioning. Despite that, sometimes positioning errors occur. In a typical case the correct position alternates with a distant point, probably due to changing between different positioning methods within the fused location provider. An example of such a problem in shown in \figref{example-positioning-problem}.
\begin{figure}
\begin{center}
\includegraphics[width=0.4\columnwidth]{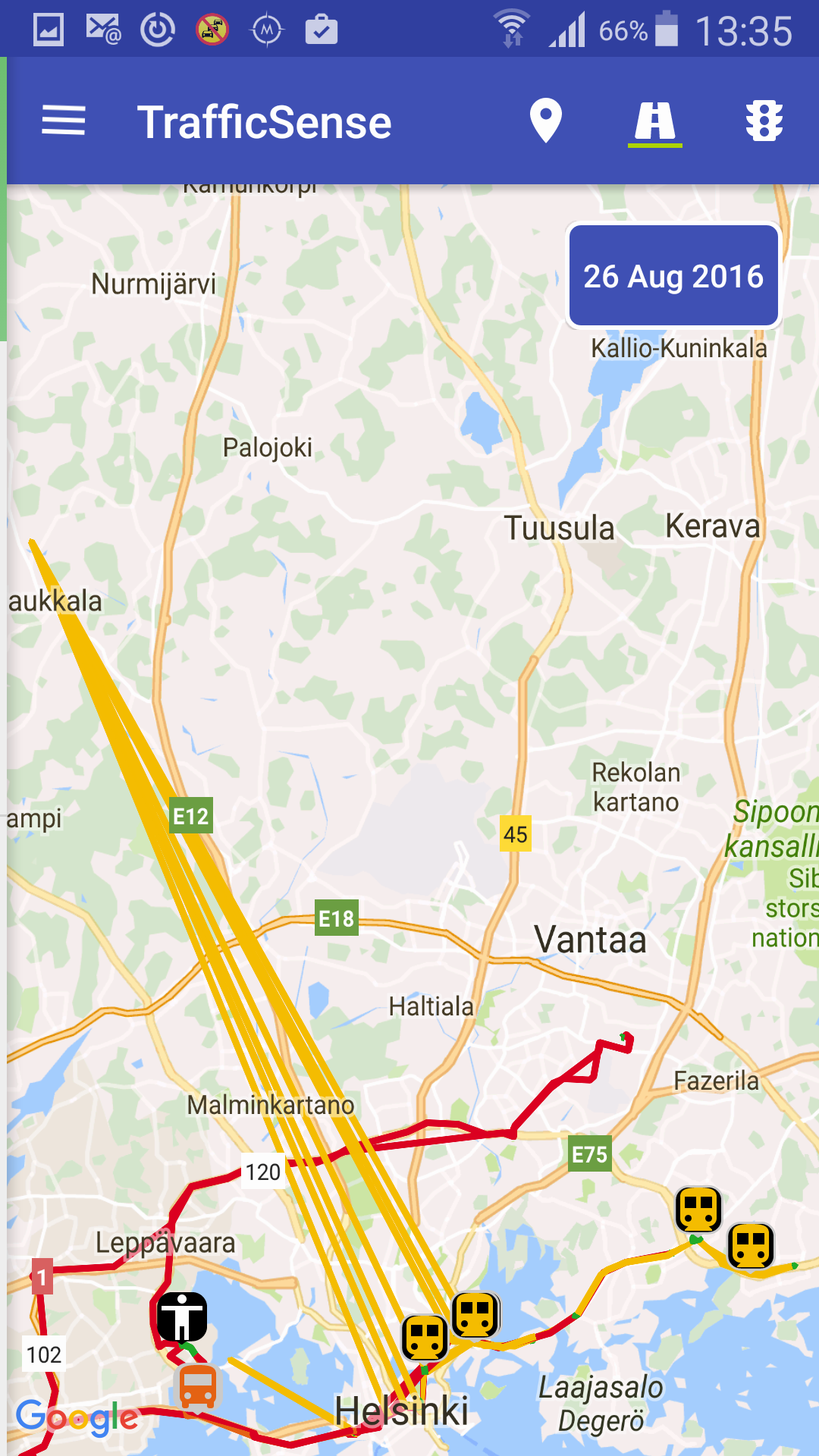}
\caption{An example of a positioning problem, where the correct position fix and an incorrect distant point are alternating (orange lines, during a subway trip). Map data \copyright 2016 Google.}
\figlabel{example-positioning-problem}
\end{center}
\end{figure}
\subsectionlabeled{device-data-filtered}{Filtered device data}
The dataset also includes a table of filtered \texttt{device\_data}. The filtering operation serves two main purposes:
\begin{itemize}
\item Remove periods during which the terminal was not substantially moving.
\item Find the most likely activity for each remaining data point.
\end{itemize}
Filtered device data is included in the published data set, because some of our recognition algorithms use it. A new candidate solution is welcome to base itself on the more complete \texttt{device\_data} instead, and implement other filtering approaches. The activity filtering algorithm has a clear impact on recognition results, as can be seen from the data shown in \sectionref{transit-live}. The following parameters are included in \texttt{device-data-filtered} (corresponding descriptions are the same as in \tabref{device-data}):
\begin{enumerate}
\item \texttt{time} (timestamp no tz)
\item \texttt{device\_id} (integer)
\item \texttt{lat} (double, latitude)
\item \texttt{lng} (double, longitude)
\item \texttt{activity} (enum value of the winning activity)
\end{enumerate}
The table contains 5975 points. The CSV-version is sorted by \texttt{time} and \texttt{device\_id}.
\subsectionlabeled{device-models}{List of device models}
A list of the models of the eight smartphones used by the test participants is provided as a separate table. It is included, because some differences were observed in \eg activity recognition performance between different devices. The table contains the following columns:
\begin{enumerate}
\item \texttt{device\_id} (integer, same as in \tabref{device-data})
\item \texttt{model} (string name of the model)
\end{enumerate}
\sectionlabeled{pubtrans-infra}{Public transport infrastructure information}
Details of the information sourced from public transport infrastructure providers is described in this section.
\subsectionlabeled{transit-live}{Live positions of public transport vehicles}
The live positions of the public transport fleet were obtained from HRT and sampled at 30 second intervals. The columns recorded into the dataset are shown in \tabref{transit-live-table}. The time period was restricted to the time of the trial. The maximum and minimum coordinates recorded by the participants were checked and the live vehicle position data was filtered (as a rectangle) to include only the area surrounding the locations sampled from the test participants. The resulting area is shown in \figref{transit-live-boundaries}. The table length is 229,451 entries.
\begin{table}[htp]
\caption{Live positions of public transport vehicles (tz = timezone). All data as provided by the transit live data API.}
\begin{center}
\begin{tabular}{|l|l|l|}
\hline
\textbf{Label} & \textbf{Type} & \textbf{Description}\\
\hline
\texttt{time} & timestamp (no tz) & The time the position was recorded.\\
\hline
\texttt{lat} & double & Vehicle location latitude, in WGS84.\\
\hline
\texttt{lng} & double & Vehicle location longitude, in WGS84.\\
\hline
\texttt{line\_type} & enum & One of: \texttt{SUBWAY} / \texttt{BUS} / \texttt{TRAM} / \texttt{TRAIN} / \texttt{FERRY}.\\
\hline
\texttt{line\_name} & string & Identifier for the public transport line, \eg 7A, 102T.\\
\hline
\texttt{vehicle\_ref} & string & Distinguish between different vehicles with the same\\
& & \texttt{line\_name} in traffic at the same time.\\
\hline
\end{tabular}
\end{center}
\tablabel{transit-live-table}
\end{table}
\begin{figure}
\begin{center}
\includegraphics[width=0.4\columnwidth]{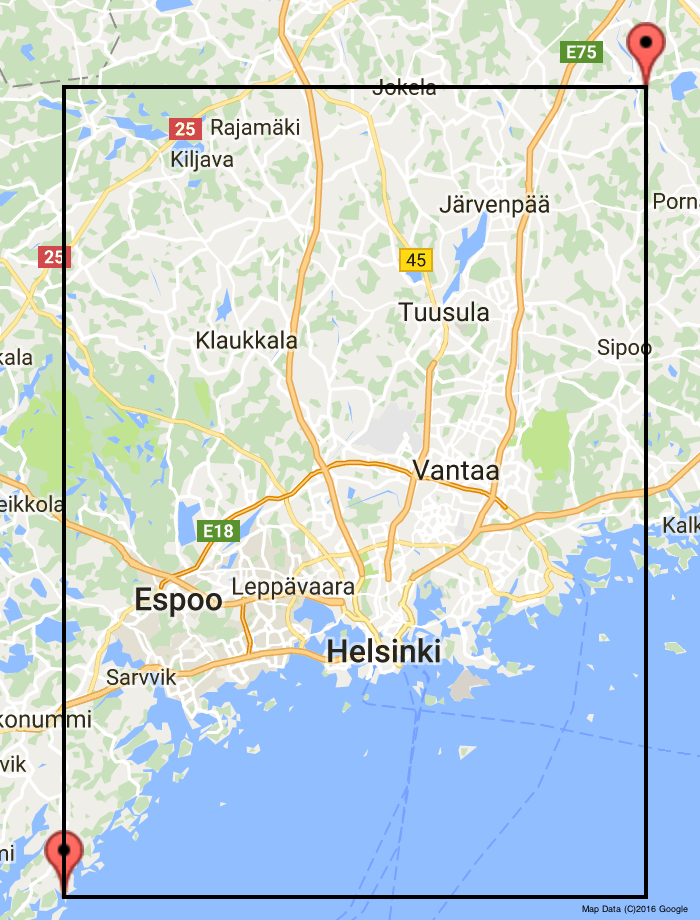}
\caption{Transit live data boundaries. The area inside the rectangle is included in the dataset. Map Data \copyright 2016 Google.}
\figlabel{transit-live-boundaries}
\end{center}
\end{figure}
\subsectionlabeled{static-timetables}{Static timetables}
The static timetables from HRT are not included in the data repository, but the file is available through the following link: \url{http://dev.hsl.fi/gtfs/hsl_20160825T125101Z.zip}. It can be used \eg with an \emph{OpenTripPlanner\footnote{\url{http://www.opentripplanner.org/}}} (OTP) server to query for trips with matching start and end locations and start times. A more precise description for the specific task of querying timetable information can be found in the documentation for the digitransit\footnote{\url{https://www.digitransit.fi/en/developers/services-and-apis/1-routing-api/itinerary-planning/}} API. The data includes the static timetables for all the public transport vehicles (also local trains) used by the study participants.
\subsectionlabeled{train-history}{Train stop times}
The dataset also includes information about the recorded stop times of local trains at stations on the day of the study. The information\footnote{\url{https://github.com/aalto-trafficsense/public-transport-dataset/tree/master/trains-json}} includes a JSON-format \texttt{junat} object, including the stopping times of trains at each station as described (in Finnish) at \url{http://rata.digitraffic.fi/api/v1/doc/index.html#Junavastaus}. Information on the referenced stations, including their locations, can be obtained as \url{http://rata.digitraffic.fi/api/v1/metadata/stations}. The description of the stations ``Liikennepaikat'' format is available (in Finnish) at \url{http://rata.digitraffic.fi/api/v1/doc/index.html#Liikennepaikkavastaus}.
\sectionlabeled{limitations}{Data coverage and limitations}
No local trains and not all the buses are included in the live transit data. Travelled \texttt{line\_name} specifiers (appearing in the manually logged data) \textbf{found} in \texttt{transit\_live} are:
\begin{itemize}
\item Trams: 2, 3, 7A, 8, 9
\item Buses: 16, 67, 72, 550, 560
\item Subways (line\_name in manual-log varies and cannot be used for comparison)
\end{itemize}
Travelled \texttt{line\_name} specifiers \textbf{not found} in \texttt{transit-live} are:
\begin{itemize}
\item Trains: E, I, K, P, U
\item Buses: Espoo18\footnote{Bus \texttt{line\_name} specifiers $<$100 can be re-used by the adjoining cities of Helsinki, Espoo and Vantaa. Bus 18 from Helsinki is included in \texttt{transit\_live}, but the test participant used bus 18 of Espoo.}, 95, 102T, 103T, 105, 110, 132T, 154, 156
\end{itemize}

In terms of recorded trips, the 10 bus trips shown in \tabref{logged-trips-not-found-in-live} can therefore definitely not be found in \texttt{transit\_live}. The 28 trips in \tabref{live-trips-recognised} can be confirmed as being discoverable and identifiable using transit live data, as they have been found with our algorithm.
\begin{table}[htp]
\caption{Manually logged bus trips, which are \textbf{not} available in live transit data.}
\begin{center}
\begin{tabular}{|l|l|l|}
\hline
\textbf{device\_id} & \textbf{Departure time} & \textbf{line\_name}\\
\hline
1 & 09:07:00 & 154\\
\hline
5 & 09:08:00 & 110\\
\hline
5 & 09:16:00 & 18 (Espoo)\\
\hline
3 & 09:59:32 & 132T\\
\hline
 2 & 10:38:51 & 103T\\
\hline
6 & 13:26:00 & 95\\
\hline
2 & 13:34:35 & 103T\\
\hline
3 & 14:21:08 & 105\\
\hline
3 & 15:26:39 & 102T\\
\hline
1 & 15:59:00 & 156\\
\hline
\end{tabular}
\end{center}
\tablabel{logged-trips-not-found-in-live}
\end{table}
\begin{table}[htp]
\caption{Manually logged trips correctly recognised from live data (28) using our algorithms (logged trips matching multiple segments have multiple rows).}
\scriptsize
\begin{center}
\begin{tabular}{|l|l|l|l|l|l|l|l|l|l|l|}
\hline
\textbf{dev\_} & \textbf{log\_} & \textbf{log\_end} & \textbf{log\_type} & \textbf{log\_} & \textbf{id} & \textbf{segm\_} & \textbf{segm\_} & \textbf{activity} & \textbf{recd\_} & \textbf{recd\_}\\
\textbf{id} & \textbf{start} & & & \textbf{name} & & \textbf{start} & \textbf{end} & & \textbf{type} & \textbf{name}\\
\hline
1 & 09:41:00 & 09:44:00 & \texttt{TRAM} & 3 & 6 & 09:40:58 & 09:44:59 & \texttt{IN\_VEHICLE} & \texttt{TRAM} & 3\\
\hline
1 & 10:06:00 & 10:19:00 & \texttt{TRAM} & 7A & 12 & 10:07:24 & 10:13:23 & \texttt{IN\_VEHICLE} & \texttt{TRAM} & 7A\\
\hline
1 & 10:06:00 & 10:19:00 & \texttt{TRAM} & 7A & 13 & 10:13:34 & 10:14:56 & \texttt{ON\_BICYCLE} & &\\
\hline
1 & 10:06:00 & 10:19:00 & \texttt{TRAM} & 7A & 14 & 10:15:37 & 10:20:11 & \texttt{IN\_VEHICLE} & \texttt{TRAM} & 7A\\
\hline
1 & 10:46:00 & 10:52:00 & \texttt{SUBWAY} & V & 19 & 10:43:41 & 10:52:45 & \texttt{IN\_VEHICLE} & \texttt{SUBWAY} & V\\
\hline
1 & 11:01:00 & 11:06:00 & \texttt{SUBWAY} & & 21 & 11:01:12 & 11:09:28 & \texttt{IN\_VEHICLE} & \texttt{SUBWAY} & V\\
\hline
1 & 11:15:00 & 11:19:00 & \texttt{SUBWAY} & M & 22 & 11:16:07 & 11:20:23 & \texttt{IN\_VEHICLE} & \texttt{SUBWAY} & M\\
\hline
1 & 13:14:00 & 13:30:00 & \texttt{SUBWAY} & & 29 & 13:13:31 & 13:30:27 & \texttt{IN\_VEHICLE} & \texttt{SUBWAY} & V\\
\hline
1 & 14:18:00 & 14:30:00 & \texttt{SUBWAY} & & 34 & 14:17:32 & 14:22:30 & \texttt{IN\_VEHICLE} & \texttt{SUBWAY} & V\\
\hline
1 & 14:18:00 & 14:30:00 & \texttt{SUBWAY} & & 35 & 14:22:41 & 14:26:37 & \texttt{ON\_BICYCLE} & &\\
\hline
1 & 14:18:00 & 14:30:00 & \texttt{SUBWAY} & & 36 & 14:27:18 & 14:31:15 & \texttt{IN\_VEHICLE} & \texttt{SUBWAY} & V\\
\hline
1 & 14:38:00 & 14:51:00 & \texttt{SUBWAY} & & 37 & 14:31:25 & 14:39:23 & \texttt{WALKING} & &\\
\hline
1 & 14:38:00 & 14:51:00 & \texttt{SUBWAY} & & 38 & 14:39:54 & 14:51:15 & \texttt{IN\_VEHICLE} & \texttt{SUBWAY} & V\\
\hline
1 & 14:57:00 & 15:05:00 & \texttt{SUBWAY} & & 40 & 14:56:47 & 15:06:45 & \texttt{IN\_VEHICLE} & \texttt{SUBWAY} & V\\
\hline
1 & 15:36:00 & 15:55:00 & \texttt{SUBWAY} & & 44 & 15:36:27 & 15:42:17 & \texttt{IN\_VEHICLE} & \texttt{SUBWAY} & M\\
\hline
1 & 15:36:00 & 15:55:00 & \texttt{SUBWAY} & & 45 & 15:42:27 & 15:48:05 & \texttt{ON\_BICYCLE} & &\\
\hline
1 & 15:36:00 & 15:55:00 & \texttt{SUBWAY} & & 46 & 15:50:21 & 16:18:46 & \texttt{IN\_VEHICLE} & &\\
\hline
2 & 13:14:34 & 13:30:03 & \texttt{SUBWAY} & To west & 54 & 13:14:01 & 13:31:34 & \texttt{IN\_VEHICLE} & \texttt{SUBWAY} & V\\
\hline
3 & 10:59:27 & 11:10:08 & \texttt{SUBWAY} & To east & 74 & 10:59:43 & 11:09:46 & \texttt{IN\_VEHICLE} & \texttt{SUBWAY} & V\\
\hline
3 & 11:27:06 & 11:34:11 & \texttt{SUBWAY} & To west & 76 & 11:28:05 & 11:34:36 & \texttt{IN\_VEHICLE} & \texttt{SUBWAY} & V\\
\hline
3 & 13:14:14 & 13:24:28 & \texttt{SUBWAY} & To west & 81 & 13:15:14 & 13:23:37 & \texttt{IN\_VEHICLE} & \texttt{SUBWAY} & V\\
\hline
4 & 10:17:00 & 10:28:00 & \texttt{TRAM} & 7A & 105 & 10:14:27 & 10:38:15 & \texttt{IN\_VEHICLE} & \texttt{TRAM} & 7A\\
\hline
4 & 10:47:00 & 10:53:00 & \texttt{SUBWAY} & to east & 107 & 10:45:59 & 10:53:39 & \texttt{IN\_VEHICLE} & \texttt{SUBWAY} & V\\
\hline
4 & 13:21:50 & 13:31:00 & \texttt{SUBWAY} & to west & 116 & 13:21:20 & 13:31:57 & \texttt{IN\_VEHICLE} & \texttt{SUBWAY} & V\\
\hline
4 & 13:44:00 & 13:47:00 & \texttt{SUBWAY} & to west & 118 & 13:45:12 & 13:47:30 & \texttt{IN\_VEHICLE} & \texttt{SUBWAY} & V\\
\hline
4 & 14:28:00 & 14:31:00 & \texttt{TRAM} & 9 & 122 & 14:26:04 & 14:31:10 & \texttt{IN\_VEHICLE} & \texttt{TRAM} & 9\\
\hline
5 & 10:52:00 & 11:02:00 & \texttt{BUS} & 16 & 136 & 10:52:30 & 11:03:28 & \texttt{IN\_VEHICLE} & \texttt{BUS} & 16\\
\hline
5 & 14:40:00 & 14:50:00 & \texttt{SUBWAY} & R & 149 & 14:40:46 & 14:50:51 & \texttt{IN\_VEHICLE} & \texttt{SUBWAY} & V\\
\hline
5 & 15:10:00 & 15:20:00 & \texttt{TRAM} & 9 & 151 & 15:08:38 & 15:24:01 & \texttt{IN\_VEHICLE} & \texttt{TRAM} & 9\\
\hline
6 & 14:01:00 & 14:17:00 & \texttt{BUS} & 560 & 162 & 13:59:21 & 14:25:50 & \texttt{IN\_VEHICLE} & \texttt{BUS} & 560\\
\hline
8 & 10:18:00 & 10:23:00 & \texttt{TRAM} & 9 & 174 & 10:17:00 & 10:23:55 & \texttt{IN\_VEHICLE} & \texttt{TRAM} & 9\\
\hline
8 & 10:44:00 & 10:53:00 & \texttt{TRAM} & 9 & 178 & 10:44:22 & 10:54:44 & \texttt{IN\_VEHICLE} & \texttt{TRAM} & 9\\
\hline
8 & 11:08:00 & 11:13:00 & \texttt{BUS} & 72 & 180 & 11:07:46 & 11:14:48 & \texttt{IN\_VEHICLE} & \texttt{BUS} & 72\\
\hline
8 & 11:25:00 & 11:41:00 & \texttt{SUBWAY} & & 181 & 11:26:23 & 11:42:24 & \texttt{IN\_VEHICLE} & \texttt{SUBWAY} & M\\
\hline
8 & 13:49:00 & 14:04:00 & \texttt{BUS} & 550 & 187 & 13:48:41 & 14:04:30 & \texttt{IN\_VEHICLE} & \texttt{BUS} & 550\\
\hline
\end{tabular}
\end{center}
\tablabel{live-trips-recognised}
\end{table}

The 38 trips in \tabref{live-trips-with-overlapping-in-vehicle} have an overlapping \texttt{IN\_VEHICLE} (sometimes also an \texttt{ON\_BICYCLE}) segment in \texttt{device\_data\_filtered}. Therefore these trips should be recognisable if the particular vehicle exists in live, but as shown in the \texttt{recd\_type} and \texttt{recd\_line} columns, they have not been properly recognised by the current algorithms. Logged trips matching multiple recorded segments are listed multiple times in the table. For performance gains in recognising trips using live data, these are expected to be the best candidates to look at. 
\begin{table}[htp]
\caption{Manually logged trips having a corresponding \texttt{IN\_VEHICLE} segment in sampled data (38), but not correctly recognised from \texttt{transit\_live} by the current algorithms.}
\scriptsize
\begin{center}
\begin{tabular}{|l|l|l|l|l|l|l|l|l|l|l|}
\hline
\textbf{dev\_} & \textbf{log\_} & \textbf{log\_end} & \textbf{log\_type} & \textbf{log\_} & \textbf{id} & \textbf{segm\_} & \textbf{segm\_} & \textbf{activity} & \textbf{recd\_} & \textbf{recd\_}\\
\textbf{id} & \textbf{start} & & & \textbf{name} & & \textbf{start} & \textbf{end} & & \textbf{type} & \textbf{name}\\
\hline
1 & 09:31:00 & 09:35:00 & \texttt{SUBWAY} & V & 4 & 09:31:44 & 09:36:37 & \texttt{IN\_VEHICLE} & &\\
\hline
1 & 09:48:00 & 09:51:00 & \texttt{SUBWAY} & & 8 & 09:49:19 & 09:54:48 & \texttt{IN\_VEHICLE} & &\\
\hline
1 & 10:28:00 & 10:35:00 & \texttt{SUBWAY} & & 17 & 10:28:00 & 10:37:33 & \texttt{IN\_VEHICLE} & &\\
\hline
1 & 11:42:00 & 11:43:00 & \texttt{SUBWAY} & M & 24 & 11:40:36 & 11:46:04 & \texttt{IN\_VEHICLE} & &\\
\hline
1 & 11:56:00 & 12:01:00 & \texttt{SUBWAY} & & 26 & 11:55:14 & 12:02:53 & \texttt{IN\_VEHICLE} & &\\
\hline
1 & 13:43:00 & 13:52:00 & \texttt{TRAM} & 9 & 31 & 13:41:40 & 13:46:33 & \texttt{ON\_BICYCLE} & &\\
\hline
1 & 13:43:00 & 13:52:00 & \texttt{TRAM} & 9 & 32 & 13:46:54 & 14:05:12 & \texttt{IN\_VEHICLE} & &\\
\hline
1 & 13:58:00 & 14:04:00 & \texttt{SUBWAY} & & 32 & 13:46:54 & 14:05:12 & \texttt{IN\_VEHICLE} & &\\
\hline
1 & 15:24:00 & 15:26:00 & \texttt{SUBWAY} & & 42 & 15:24:38 & 15:28:24 & \texttt{IN\_VEHICLE} & &\\
\hline
2 & 11:04:00 & 11:06:33 & \texttt{SUBWAY} & To east & 50 & 11:04:13 & 11:12:20 & \texttt{IN\_VEHICLE} & &\\
\hline
2 & 11:26:18 & 11:50:00 & \texttt{SUBWAY} & To east & 51 & 11:22:41 & 11:58:31 & \texttt{IN\_VEHICLE} & &\\
\hline
2 & 11:50:44 & 11:58:40 & \texttt{SUBWAY} & To west & 51 & 11:22:41 & 11:58:31 & \texttt{IN\_VEHICLE} & &\\
\hline
3 & 10:21:21 & 10:23:01 & \texttt{SUBWAY} & To east & 70 & 10:17:28 & 10:25:54 & \texttt{IN\_VEHICLE} & &\\
\hline
3 & 10:37:47 & 10:47:59 & \texttt{SUBWAY} & To east & 72 & 10:36:03 & 10:48:14 & \texttt{IN\_VEHICLE} & &\\
\hline
3 & 11:56:11 & 11:58:43 & \texttt{SUBWAY} & To west & 78 & 11:57:56 & 12:01:18 & \texttt{IN\_VEHICLE} & &\\
\hline
3 & 13:35:13 & 13:41:37 & \texttt{TRAM} & 7A & 83 & 13:28:48 & 13:41:28 & \texttt{IN\_VEHICLE} & &\\
\hline
3 & 13:45:37 & 13:47:12 & \texttt{SUBWAY} & To west & 85 & 13:45:42 & 13:48:24 & \texttt{IN\_VEHICLE} & &\\
\hline
3 & 14:07:59 & 14:09:23 & \texttt{SUBWAY} & To west & 87 & 13:59:48 & 14:14:14 & \texttt{IN\_VEHICLE} & &\\
\hline
3 & 14:11:30 & 14:13:34 & \texttt{SUBWAY} & To west & 87 & 13:59:48 & 14:14:14 & \texttt{IN\_VEHICLE} & &\\
\hline
3 & 14:42:40 & 14:52:31 & \texttt{SUBWAY} & To east & 90 & 14:39:47 & 14:52:22 & \texttt{IN\_VEHICLE} & &\\
\hline
3 & 15:02:03 & 15:13:57 & \texttt{SUBWAY} & To west & 92 & 15:01:12 & 15:16:46 & \texttt{IN\_VEHICLE} & &\\
\hline
4 & 09:39:00 & 09:45:00 & \texttt{TRAM} & 9 & 97 & 09:38:15 & 09:41:02 & \texttt{IN\_VEHICLE} & \texttt{TRAM} & 3\\
\hline
4 & 09:39:00 & 09:45:00 & \texttt{TRAM} & 9 & 98 & 09:41:12 & 09:42:25 & \texttt{ON\_BICYCLE} & &\\
\hline
4 & 09:39:00 & 09:45:00 & \texttt{TRAM} & 9 & 99 & 09:43:08 & 09:45:15 & \texttt{IN\_VEHICLE} & \texttt{TRAM} & 1\\
\hline
4 & 10:35:00 & 10:37:00 & \texttt{SUBWAY} & to east & 105 & 10:14:27 & 10:38:15 & \texttt{IN\_VEHICLE} & \texttt{TRAM} & 7A\\
\hline
4 & 11:04:00 & 11:13:00 & \texttt{SUBWAY} & to east & 109 & 11:04:55 & 11:14:10 & \texttt{IN\_VEHICLE} & &\\
\hline
4 & 11:26:00 & 11:28:00 & \texttt{SUBWAY} & to west & 111 & 11:23:59 & 11:31:36 & \texttt{IN\_VEHICLE} & &\\
\hline
4 & 11:36:00 & 11:38:00 & \texttt{SUBWAY} & to east & 113 & 11:36:32 & 11:46:49 & \texttt{IN\_VEHICLE} & &\\
\hline
4 & 11:45:00 & 11:47:00 & \texttt{SUBWAY} & to east & 113 & 11:36:32 & 11:46:49 & \texttt{IN\_VEHICLE} & &\\
\hline
4 & 13:57:00 & 14:01:00 & \texttt{SUBWAY} & to west & 120 & 13:57:00 & 14:02:39 & \texttt{IN\_VEHICLE} & &\\ 
\hline
4 & 15:15:00 & 15:20:00 & \texttt{SUBWAY} & to east & 125 & 15:15:29 & 15:20:13 & \texttt{IN\_VEHICLE} & &\\
\hline
5 & 09:58:00 & 10:08:00 & \texttt{TRAM} & 7A & 133 & 09:58:03 & 10:14:13 & \texttt{IN\_VEHICLE} & &\\
\hline
5 & 10:12:00 & 10:13:00 & \texttt{SUBWAY} & M & 133 & 09:58:03 & 10:14:13 & \texttt{IN\_VEHICLE} & &\\
\hline
5 & 11:11:00 & 11:14:00 & \texttt{SUBWAY} & R & 138 & 11:09:19 & 11:14:28 & \texttt{IN\_VEHICLE} & &\\
\hline
5 & 11:18:00 & 11:20:00 & \texttt{BUS} & 67 & 140 & 11:18:23 & 11:20:53 & \texttt{IN\_VEHICLE} & &\\
\hline
5 & 11:38:00 & 11:52:00 & \texttt{SUBWAY} & M & 142 & 11:38:53 & 11:50:59 & \texttt{IN\_VEHICLE} & &\\
\hline
5 & 15:36:00 & 15:39:00 & \texttt{TRAM} & 9 & 153 & 15:37:43 & 15:40:10 & \texttt{IN\_VEHICLE} & &\\
\hline
6 & 11:38:00 & 11:52:00 & \texttt{SUBWAY} & & 156 & 11:45:26 & 11:50:26 & \texttt{IN\_VEHICLE} & &\\
\hline
6 & 14:19:00 & 14:25:00 & \texttt{SUBWAY} & & 162 & 13:59:21 & 14:25:50 & \texttt{IN\_VEHICLE} & BUS & 560\\
\hline
6 & 15:50:00 & 15:59:00 & \texttt{TRAM} & 9 & 165 & 15:50:58 & 15:59:24 & \texttt{IN\_VEHICLE} & &\\
\hline
8 & 11:52:00 & 11:58:00 & \texttt{SUBWAY} & & 183 & 11:52:35 & 12:00:18 & \texttt{IN\_VEHICLE} & &\\
\hline
\end{tabular}
\end{center}
\tablabel{live-trips-with-overlapping-in-vehicle}
\end{table}

Finally, the 13 trips in \tabref{live-trips-no-overlapping-in-vehicle} have no overlapping \texttt{IN\_VEHICLE} segment from the mobile device samples. It is also noted that:
\begin{itemize}
\item 3/13 trips (2 subway + 1 tram) have no overlapping mobile device samples at all
\item 10/13 trips overlap with \texttt{WALKING} (in 6/10 cases shadowing the whole trip)
\item 12/13 trips (\texttt{device\_id} 5 and 6) were carried out with the same device model
\end{itemize}
The \texttt{device\_data} sampled before and after a time segment with no data should still reveal that the mobile device has moved without recording positions, \ie it has either erroneously been in \texttt{SLEEP} state or no position fixes have been received. Looking at the timestamps and locations recorded before and after the break period with geographical displacement, a set of candidate options could still be found from either the static timetable or live fleet position data. In downtown areas with many transport alternatives errors would be likely, but in less busy areas the set of alternatives will be smaller. Faulty recognitions changing car trips to public transport would still be very likely, so this type of recognition should only be attempted in scenarios where the penalty of false positive is not very high. This type of recognition has not been attempted in the current experiment.
\begin{table}[htp]
\caption{Manually logged trips having \textbf{no} corresponding \texttt{IN\_VEHICLE} segment (13) in sampled data.}
\scriptsize
\begin{center}
\begin{tabular}{|l|l|l|l|l|l|l|l|l|}
\hline
\textbf{dev\_} & \textbf{log\_} & \textbf{log\_end} & \textbf{log\_type} & \textbf{log\_} & \textbf{id} & \textbf{segm\_} & \textbf{segm\_} & \textbf{activity}\\
\textbf{id} & \textbf{start} & & & \textbf{name} & & \textbf{start} & \textbf{end} &\\
\hline
4 & 15:53:00 & 15:55:00 & \texttt{SUBWAY} & to west & 127 & 15:41:39 & 15:58:40 & \texttt{WALKING}\\
\hline
5 & 11:30:00 & 11:31:00 & \texttt{SUBWAY} & R & 141 & 11:21:26 & 11:36:31 & \texttt{WALKING}\\
\hline
5 & 13:22:00 & 13:28:00 & \texttt{SUBWAY} & M & 144 & 13:17:08 & 13:23:38 & \texttt{WALKING}\\
\hline
5 & 13:34:00 & 13:43:00 & \texttt{SUBWAY} & R & 145 & 13:28:54 & 13:37:27 & \texttt{WALKING}\\
\hline
5 & 13:54:00 & 14:02:00 & \texttt{SUBWAY} & V & & & &\\
\hline
5 & 14:20:00 & 14:30:00 & \texttt{SUBWAY} & R & 148 & 14:02:42 & 14:39:19 & \texttt{WALKING}\\
\hline
5 & 15:36:00 & 15:39:00 & \texttt{TRAM} & 9 & 152 & 15:34:32 & 15:37:11 & \texttt{WALKING}\\
\hline
5 & 15:51:00 & 15:52:00 & \texttt{TRAM} & 8 & & & &\\
\hline
5 & 15:56:00 & 15:57:00 & \texttt{SUBWAY} & M & & & &\\
\hline
6 & 13:47:00 & 13:49:00 & \texttt{SUBWAY} & & 161 & 13:44:23 & 13:57:30 & \texttt{WALKING}\\
\hline
6 & 14:29:00 & 14:30:00 & \texttt{SUBWAY} & & 163 & 14:26:10 & 15:03:54 & \texttt{WALKING}\\
\hline
6 & 15:02:00 & 15:20:00 & \texttt{SUBWAY} & & 163 & 14:26:10 & 15:03:54 & \texttt{WALKING}\\
\hline
6 & 15:41:00 & 15:43:00 & \texttt{TRAM} & 2 & 164 & 15:27:41 & 15:50:48 & \texttt{WALKING}\\
\hline
\end{tabular}
\end{center}
\tablabel{live-trips-no-overlapping-in-vehicle}
\end{table}
\sectionlabeled{recognition-methods}{Methods of automatic recognition}
The collected live public transport vehicle locations and static timetables, in
conjunction with the user location traces, were used to automatically recognise
public transport trips taken by users. Trip legs consisting, after filtering, of a
continuous sequence of \texttt{IN\_VEHICLE} activity were matched against the vehicle position
traces and timetable data. The train stop times were not used.
\subsectionlabeled{live-vehicle-location}{Live vehicle location}
The vehicle location data from \tabref{transit-live-table} is compared with the user locations
collected by the personal mobile devices as described in \sectionref{device-data}. Potential issues in matching include:
\begin{itemize}
\item Missing vehicle or user data points
\item Inaccurate location points
\item Clock differences
\item Distance between user and vehicle location sensor in longer vehicles
\item Distance between location samples at higher vehicle velocities
\item False matches to other public transport vehicles
\item False positives where a car trip takes place near public transport vehicles
\item Intermittent changes of the line name label on some vehicles in the live data
\end{itemize}
For performance reasons the number of user location samples used for matching
a trip leg was limited to 40. To counteract the location accuracy and vehicle length issues, a distance limit
of 100 metres was used for collecting vehicle matches. A greater limit may
cause more false positive matches to appear.

With the sample rate of thirty seconds in the collected vehicle positions, a
vehicle traveling at 80 km/h would produce samples every 667 metres, much in
excess of the above mentioned distance limit. This could cause false negatives
with the user location samples falling in between the vehicle points in such a
way that they are not matched. To prevent this, the position sequence of each
vehicle is processed into linestrings, and user point distances calculated
against those line geometries.

The vehicle location points for composing the linestrings for comparison are
collected in a $\pm60$-second window around the timestamp of each user point
sample. This allows for some clock difference, and sampling time difference.

For a vehicle to be accepted as a possible match, its linestring must be within
the 100 metre distance limit for a minimum of 75\% of the user location
samples.

Each match within the permitted 100 metres' distance accumulates a score of
$100 - d$ when the distance is $d$ metres. The vehicle with the highest score
wins.

The line type and name are set according to the matched vehicle. In case of the
vehicle having multiple line names in the matches, the most frequently
occurring name is used. On some lines, a vehicle can intentionally change line
name when passing a certain stop. Also, some vehicles in the data set change
line name intermittently to false values.

In addition to the method described above (subsequently referred to as ``New
live''), a prior implementation (``Old live'') was evaluated. In the older
implementation, four user trace sample points were used, and matched against
vehicle location points in the surrounding time window. With the vehicle
location sampling interval of 30 seconds, and the 100 metre distance criterion,
this would be expected to cause otherwise optimal user trace point samples to
potentially fall outside the matching radius once the vehicle speed exceeds
$2 \times 100\mathrm{m} / 30\mathrm{s} \approx 6.67 \mathrm{m/s} \approx
24 \mathrm{km/h}$.

For trams in Helsinki having an average speed of 14.7 km/h
(2013--2014)\footnote{\url{https://www.hsl.fi/sites/default/files/uploads/13_2015_raitioliikenteen_linjastosuunnitelma_netti.pdf}},
with dense stops the new and old methods should produce similar results. More
differences can be expected on the subway, and on bus routes with highway
segments, where speeds are higher and stops more sparse.
\subsectionlabeled{matching-with-static-timetables}{Matching with static timetables}
%
Comparison with static timetables is based on searching for public transport plans of past trips based on the sampled \texttt{IN\_VEHICLE} segments.
Firstly, a trip-plan query is sent to the OTP journey planner interface based on each candidate leg's start-time, first point (origin) and last point (destination). Secondly, the resulting PT plans are compared to the leg's start-time and end-time as well as the user location trace to identify whether the leg represents a PT ride. 
Potential issues in matching include:
\begin{itemize}
	\item Missing or inaccurate user location points
	\item Inaccuracy in activity determination and filtered transition points
	\item Clock differences
	\item False positives when the route and timing of a car trip are similar to a segment of a public transport line
\end{itemize}

\tabref{mass-transit-match-table-a} explains the constants and \tabref{mass-transit-match-table-b} the variables and formulas used in the process. \figref{mass-transit-match} illustrates an example public transport trip and highlights the following parameters showing possible extra parts of the planned trip compared to the original sampled trip: $tWb, tWe, tPTb, tPTe$, where $t$ indicates "time", $W$ indicates "walking" and $PT$ indicates "public transport". The $b$ denotes "at the beginning of a trip or leg", and $e$ denotes "at the end of a trip or leg".

\begin{table}[htp]
	\caption{Constants used in our query to the Open Trip Planner, validation and matching of the resulting trip plan with the recorded trip.}
	\scriptsize
	\begin{center}
	\tablabel{mass-transit-match-table-a}	
	\begin{tabular}{|l|l|l|}
		\hline
		\textbf{Parameter} & \textbf{Definition}                                                                                                                                                                                                                                        & \textbf{Value} \\ \hline
		$dEmax$ & \begin{tabular}[c]{@{}l@{}}Maximum acceptable distance inaccuracy as a result of displacement \\ or delay in detecting the point of mode transition. This value denotes \\ an offset in start point and/or end point of filtered IN\_VEHICLE leg.\end{tabular} & 500 m               \\ \hline
		$vW$                 & Walking speed.                                                                                                                                                                                                                                            & \textless= 1.34 m/s \\ \hline
		$vPT$                & Speed of a PT vehicle.                                                                                                                                                                                                                                 & \textgreater= 3 m/s \\ \hline
		$tEPT$               & Deviation of the PT vehicle from schedule.                                                                                                                                                                                                                  & \textless= 3 min    \\ \hline
	\end{tabular}
	\end{center}	
\end{table}

\begin{table}[htp]
	\caption{Variables and formulas used in our query to the Open Trip Planner, validation and matching of the resulting trip plan with recorded trip.}
	\scriptsize
	\tablabel{mass-transit-match-table-b}
	\begin{center}
	\begin{tabular}{|l|l|l|l|}
		\hline
		\textbf{Parameter} & \textbf{Definition} & \textbf{Formula}                                                                                                                               & \textbf{Value} \\ \hline
		$tV$                 & \begin{tabular}[c]{@{}l@{}}Duration of the sampled IN\_VEHICLE trip leg.\end{tabular}                                                     &                                                                                                                                                         &                     \\ \hline
		$t$                  & Duration of the planned trip.                                                                                                                  &       & From OTP            \\ \hline
		$dEb$                & \begin{tabular}[c]{@{}l@{}}Distance offset (error)\\ 			in detecting the starting point of a leg.\end{tabular}                                 &                                                                                                                                                         & \textless= dEmax    \\ \hline
		$dEe$                & \begin{tabular}[c]{@{}l@{}}Distance offset (error)\\ 			in detecting the end point of a leg.\end{tabular}                                      &                                                                                                                                                         & \textless= dEmax    \\ \hline
		$tWb$                & \begin{tabular}[c]{@{}l@{}}Walking duration at the beginning\\ 			of the planned trip to reach the boarding stop.\end{tabular}                     & \begin{tabular}[c]{@{}l@{}}
			$dEb/vW$\\ For	maximum acceptable inaccuracy: \\ 
			$tWb = 500/1.34 = 6.2 min$\end{tabular} & \textless= 6.2 min  \\ \hline
		$tWe$                & \begin{tabular}[c]{@{}l@{}}Walking duration at the end of the planned trip\\ 			 from the end stop to the destination.\end{tabular}                     & $dEe/vW$                                                                                                                                                  & \textless= 6.2 min  \\ \hline
		$tPT$                & \begin{tabular}[c]{@{}l@{}}Duration of the PT\\ 			leg included in the planned trip.\end{tabular}                               &                                                                                                                                                         & From OTP            \\ \hline
		$tPTb$               & \begin{tabular}[c]{@{}l@{}}Extra transit ride at the beginning of the planned \\ trip compared to the sampled IN\_VEHICLE trip leg.\end{tabular} & $dEb/vPT$                                                                                                                                                 & \textless= 2.8 min  \\ \hline
		$tPTe$               & \begin{tabular}[c]{@{}l@{}}Extra transit ride at the end of the planned trip \\compared to the sampled IN\_VEHICLE trip leg.\end{tabular}       & $dEe/vPT$                                                                                                                                                 & \textless= 2.8 min  \\ \hline
		$\Delta tPT$               & \begin{tabular}[c]{@{}l@{}}Extra transit ride of\\ 			the planned trip compared to the sampled trip.\end{tabular}                                  & \begin{tabular}[c]{@{}l@{}}$|tPT - tV| = tPTb + tPTe$\end{tabular}                                                                                   & \textless= 5.6 min  \\ \hline
		$\Delta tW$                & \begin{tabular}[c]{@{}l@{}}Extra walking in the planned\\ 			trip compared to the sampled trip\end{tabular}                                       & $tWb + tWe$                                                                                                                                               & \textless= 12.4 min \\ \hline
		$\Delta t$                 & \begin{tabular}[c]{@{}l@{}}Duration difference between the planned trip \\and the sampled IN\_VEHICLE leg.\end{tabular}                         & \begin{tabular}[c]{@{}l@{}}$\Delta tPT + \Delta tW =$\\ $tWb + tPTb + tPTe + tWe$\end{tabular}                                                                    & \textless= 18 min   \\ \hline
	\end{tabular}
	\end{center}
\end{table}

\begin{figure}
\begin{center}
	\includegraphics[width=1.0\columnwidth]{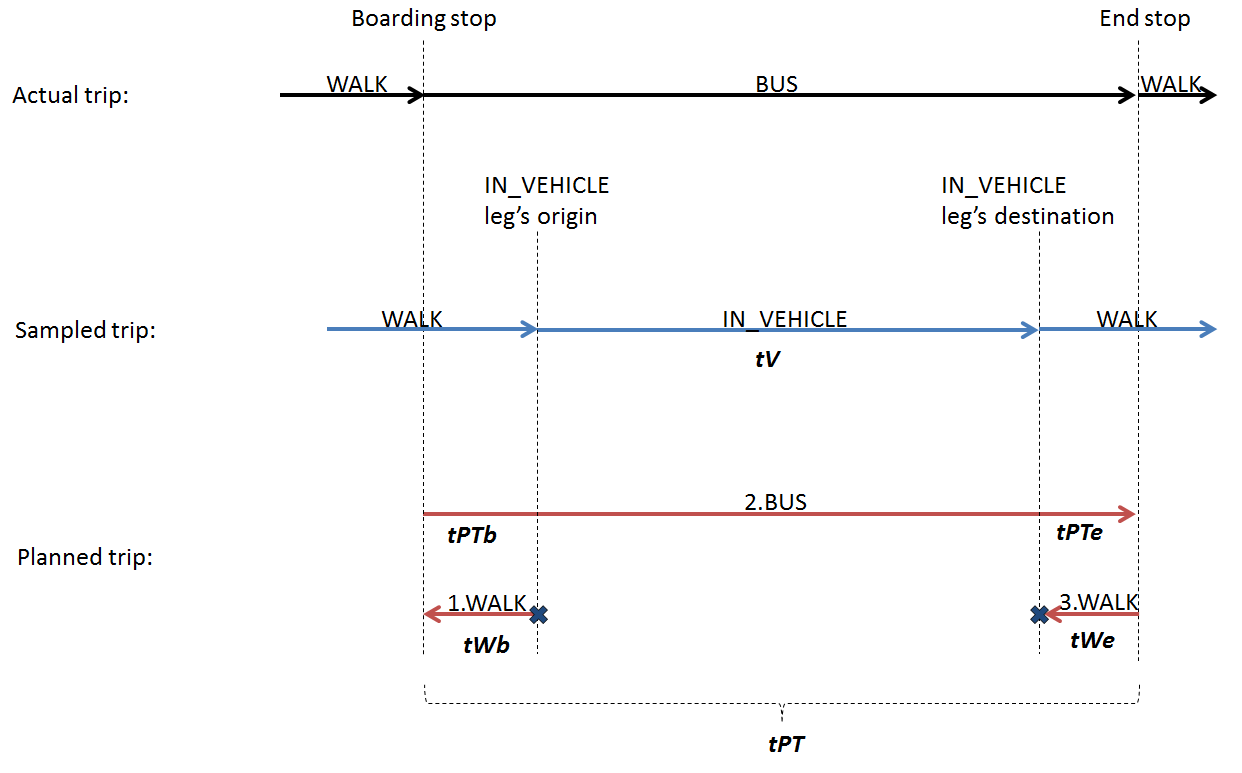}
	\caption{Matching a sampled trip with the static public transport timetable to detect whether or not the trip has been a PT trip as well as to identify the PT mode (e.g. Tram, Bus, Commuter Train).}
	\figlabel{mass-transit-match}
\end{center}
\end{figure}

\begin{figure}
\begin{center}
	\includegraphics[width=\linewidth]{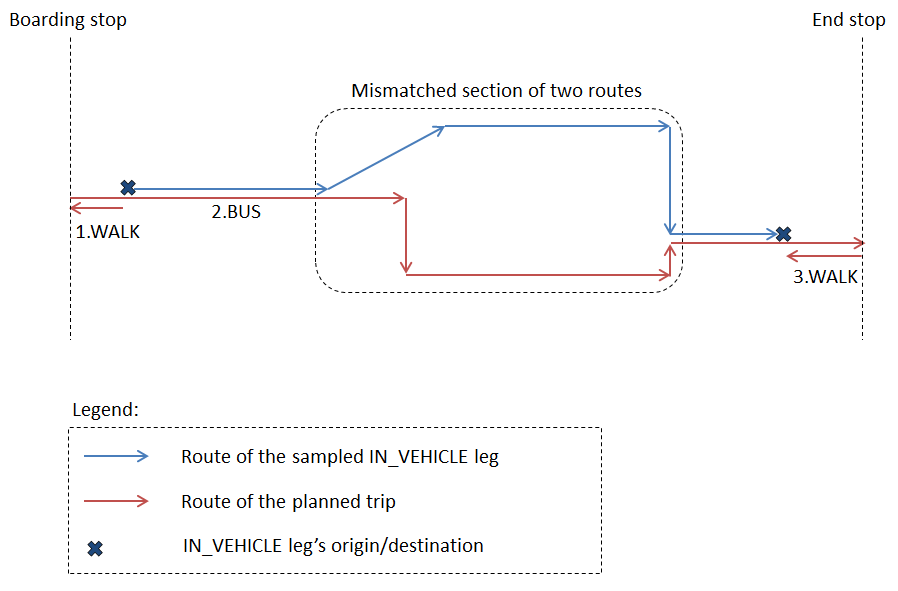}
	\caption{Matching the sampled trip route with the planned PT route. This figure illustrates an example of a mismatch between routes of planned PT transit and the sampled vehicular leg. PT vehicle of the planned trip takes a different route by partly passing different roads compared to the original sampled trip.}
	\figlabel{mass-transit-match-routes}
\end{center}
\end{figure}

The data sampled from a mobile device can often have a fair amount of inaccuracy in location and
activity detection. For this reason the filtered transition points starting and ending the
vehicular trip leg can often appear between stops during the actual mass
transit leg. Therefore almost all travel plans from the journey planner inevitably include walking sections in addition to the desired transit (vehicular) leg. The additional walking sections can appear at the beginning ($tWb$ to walk from the detected start of vehicular leg to the public transport stop for boarding) and/or at the end of the trip leg ($tWe$ to walk from the exit stop to the detected end of vehicular leg). 

%
\subsubsection{Selection of query parameters}
The origin and destination parameters of our OTP query are set equal to the leg's start and end geolocations respectively. The third query parameter, trip start-time, is set based on the leg's start-time as follows. 

The maximum allowable inaccuracy in transition points is $dEmax=500 metres$. 
Therefore, to account for inaccuracy in time and place of origin, in our query to the journey planner the earliest
permissible start-time of the trip is adjusted back by $tWb = 6.2 minutes$ to allow
500 metres of walking at a speed of $vW=1.34 m/s$ to the boarding stop. Figuratively speaking, in case of offset in recorded \texttt{IN\_VEHICLE} leg, we allow the traveller to catch the desired transit line according to actual timetable by walking to the nearest stop. If there is no delay and time and place of transition are completely accurate, the traveller would just wait 6.2 minutes at the boarding stop to take the transit line. 
Adjacent stops are assumed to be no more than one kilometre ($2 \times 500 m$) apart.

Correspondingly, to relax both the origin and the destination, a maximum of $2 \times dEmax=1,000$ metres of walking is requested for each resulting plan. The OTP query requests for three PT plans for each candidate vehicular leg.
\subsubsection{Filtering and validation of query response}
Out of the three returned plans the best match, if good enough, is chosen. Plans that match the following criteria can be \emph{discarded}:
\begin{itemize}
\item Plans having a total duration of more than $tEPT = 3 minutes$ shorter than the user-recorded vehicle leg, thereby assuming that the transit vehicle must travel closely according to schedule to avoid false positive matches.
\item Plans having a total duration of more than $\Delta dt=18 minutes$ longer than the user-recorded vehicle
leg. This difference denoted by $\Delta dt$ includes the walk times for dealing with location
inaccuracy discussed above ($\Delta tW = 6.2 min \times 2 = 12.4 min$ for the whole trip), and the time assumed for a mass transit vehicle to
travel between adjacent stops (maximum $1000 m$ for the whole trip at minimum speed of $vPT = 3  m/s$, equal to $\Delta tPT = 5.6 minutes$).
\item Plans where the duration of the included transit leg ($tPT$) differs by more than $\Delta tPT = 5.6
minutes$ from the user-recorded vehicular leg ($tV$), based on assumed
time of travel of the vehicle between adjacent sparse stops. 
\item Plans where the start of the included transit leg differs from that of the
recorded vehicular leg by more than 5.8 minutes. This value considers a public transport vehicle traveling between adjacent stops at the beginning of the trip ($tPTb = 2.8 min$, for $500 m$) plus an assumed $tEPT = 3 min$ deviation of the transit line from its schedule.
\end{itemize}
The plans returned by the journey planner interface include location point sequences of the planned trips. As illustrated in \figref{mass-transit-match-routes}, the location point sequences are also matched against the recorded user trace to verify that the route of the public transport line matches the recorded user trace. For comparison the recorded user trace is sampled at no less than 100 metre intervals, and leading and trailing points may be ignored based on the assumed inaccuracy of origin and destination. A minimum of 70\% of sample points must match a plan point within
100 metres, and no more than four adjacent sample points may fall outside 100 metres of the plan points for the plan to qualify. Out of the qualifying plans, the one with the closest start-time to that of the recorded leg wins.

Our current algorithm only validates plans containing a single vehicular leg. However, quick transfers between vehicles may not have been detected as separate vehicular legs by the
activity detection and filtering of the mobile device. Allowing trip plans with
transfers, and splitting the reference vehicular leg accordingly, could produce improved detection for such cases.
An example of the fusion of a \texttt{BUS} leg and a \texttt{SUBWAY} leg to a
single \texttt{IN\_VEHICLE} leg can be found from device 6 at 13:59 to 14:25, where a short
transfer has occurred between 14:17 and 14:19.
\sectionlabeled{recognition-results}{Recognition results}
Compared with the 103 manually logged trips, 86 \texttt{IN\_VEHICLE} segments were recognised by our filtering algorithm, 85 of them overlapping at least a part of a logged trip. 

The 28 trips recognised with correct \texttt{line\_name} from \texttt{transit\_live} using the new live algorithm were detailed in \tabref{live-trips-recognised}. The statistics for all the recognition methods are collected to \tabref{trip-matching-statistics}. The static approach yielded the highest number of matched trips, but with one false \texttt{line\_type} and two false bus \texttt{line\_name}s. Live recognition performance suffered from the absence of many buses and all trains in the data, but especially subway trip detection was significantly improved with the new live algorithm compared with the old one. Looking at the combined results but without requiring the \texttt{line\_name} to match, 60\% of bus and tram trips, 44\% of train trips and 43\% of subway trips were recognised. If the correct \texttt{line\_name} is required, bus recognition success drops to 47\%.
\begin{table}[htp]
\caption{Trip matching statistics for all recognition methods compared to the manually logged trips (``line name'' = matching line name, ``line type'' = matching line type).}
\begin{center}
\begin{tabular}{|l|l|l|l|l|l|}
\hline
& \textbf{Static} & \textbf{Old live} & \textbf{New live} & \textbf{Combined} & \textbf{Logged}\\
\hline
Bus & 8 & 3 & 4 & 9 & 15\\
\hline
Bus (line name) & 6 & 3 & 4 & 7 & 15\\
\hline
Tram & 8 & 8 & 8 & 9 & 15\\
\hline
Tram (line name) & 8 & 7 & 7 & 9 & 15\\
\hline
Train & 4 & 0 & 0 & 4 & 9\\
\hline
Train (line name) & 4 & 0 & 0 & 4 & 9\\
\hline
Subway & 19 & 9 & 17 & 25 & 58\\
\hline
Public transport & 40 & 20 & 29 & 48 & 97\\
\hline
Public transport (line type) & 39 & 20 & 29 & 47 & 97\\
\hline
\end{tabular}
\end{center}
\tablabel{trip-matching-statistics}
\end{table}
\sectionlabeled{conclusions}{Discussion and conclusions}
The referenced dataset describes the trips of seven study participants using public transportation in the Helsinki area for a day, and a reference participant using a private car. The dataset includes a manual log of the trips, automatically collected measurements from the mobile devices of the participants and the public transport infrastructure data, which was available from the public transport provider at the time of the test. The mobile device measurements include geographical position and activity estimates. The infrastructure data consists of live locations of public transport vehicles and static timetable data. The challenge is to correctly match as many public transport trips as possible using the various measurements. The results can be verified using the manual log.

The data suffers from multiple imperfections. The estimated activities from the mobile devices are not always correct, \eg sometimes a passenger is perceived to be riding a bicycle during a tram trip. Due to the power saving features of the mobile device client application and problems in positioning in trains and underground scenarios, measurements can be intermittent or completely missing. Other problems in positioning sometimes result in locations hopping between the correct position and a single distant point. Live locations were not available from all public transport vehicles at the time of the trial.

Altogether 103 trips are logged in the dataset, 97 of them carried out using public transportation. Combining matches from the static timetable and live data 60\% of bus and tram trips, and 43--44\% of train and subway trips were recognised with the correct vehicle type. Recognition of correct line names was otherwise on the same level, but for buses the recognition result dropped to 47\%. The joint combined public transport recognition reached a level of 48\% for the correct line type.

The currently achieved results are approximate, but adequate for purposes, where exact recognition of every trip is not necessary. For \eg public transport disruption information filtering the current recognition level would be sufficient, because the likelihood of a correct recognition increases fast for frequent trips on the same public transport line and the penalty for false positive recognitions is not very high. For purposes where the passenger can review past trips individually the current level of performance can be frustrating. For fare payment processing it would be unacceptable.

Improvements in activity recognition accuracy of the mobile device and better vehicle coverage of live transit data would both contribute to better recognition probability. With the current power saving algorithms the mobile device power consumption is reasonable but clearly noticeable, especially when the device is constantly in motion, causing positioning to be requested with high accuracy. Without radical improvements in battery and / or positioning technologies power consumption should be decreased rather than increased, which sets limitations to future improvements in the client sampling. One approach for testing would be to make more use of radio beacons in public transport vehicles. While in proximity of an identified radio beacon, the high accuracy positioning of the mobile device could be switched off and the system would rely on the positioning of the public transport vehicle.
\sectionlabeled{acknowledgements}{Acknowledgements}
Supported by the TrafficSense project in the Aalto Energy Efficiency Programme funded by Aalto University.

In addition to the authors the following persons have contributed to the TrafficSense software: Joonas Javanainen, Kimmo Karhu, Juho Saarela, Janne Suomalainen, Michailis Tziotis. In addition to the authors and software contributors the TrafficSense project has been participated by Mikko Heiskala, Jaakko Hollm\'en, Jani-Pekka Jokinen, Iisakki Kosonen, Esko Nuutila, Roelant Stegmann, Markku Tinnil\"a, Seppo T\"orm\"a, G\"orkem Yetik and Indre Zliobaite.
\sectionlabeled{licensing}{Licensing}
The data in the referenced github repository is licensed under the \emph{Creative Commons BY 4.0 licence\footnote{\url{http://creativecommons.org/licenses/by/4.0/}}}. The data extracted from the test group\footnote{\texttt{device\_data}, \texttt{device\_data\_filtered}, \texttt{device\_models}, \texttt{manual\_log}} is licensed by Aalto University. Train data\footnote{\texttt{commuterTrains.json}, \texttt{trainStations.json}} is obtained from Digitraffic offered by the Finnish Transport Agency. The live data on public transportation\footnote{\texttt{transit\_live}} is compiled from a service offered by Helsinki Regional Transport. Map data from the city of Helsinki\footnote{in the subway station entrance images offered in the repository} is authored by ``Helsinki, kiinteist\"oviraston kaupunkimittausosasto''.

\bibliographystyle{plain}
\bibliography{public_transport}

\end{document}